# Left Ventricular Wall Motion Estimation by Active Polynomials for Acute Myocardial Infarction Detection

Serkan Kiranyaz, Aysen Degerli, Tahir Hamid, Rashid Mazhar, Rayyan Ahmed, Rayaan Abouhasera, Morteza Zabihi, Junaid Malik, Ridha Hamila, and Moncef Gabbouj

*Abstract*— **Echocardiogram (echo) is the earliest and the primary tool for identifying regional wall motion abnormalities (RWMA) in order to diagnose myocardial infarction (MI) or commonly known as heart attack. This paper proposes a novel approach, *Active Polynomial*, which can accurately and robustly estimate the global motion of the Left Ventricular (LV) wall from any echo in a robust and accurate way. The proposed algorithm quantifies the true wall motion occurring in LV wall segments so as to assist cardiologists diagnose early signs of an acute MI. It further enables medical experts to gain an enhanced visualization capability of echo images through color-coded segments along with their "maximum motion displacement" plots helping them to better assess wall motion and LV Ejection-Fraction (LVEF). The outputs of the method can further help echo-technicians to assess and improve the quality of the echocardiogram recording. A major contribution of this study is the first public echo database collection composed by physicians at the Hamad Medical Corporation Hospital in Qatar. The so-called HMC-QU database will serve as the benchmark for the forthcoming relevant studies. The results over HMC-QU dataset show that the proposed approach can achieve high accuracy, sensitivity and precision in MI detection even though the echo quality is quite poor, and the temporal resolution is low.**

## I.    INTRODUCTION

Early detection of an acute myocardial infarction (MI) [1], [2] or in general a coronary artery disease (CAD) requires an accurate estimation of the regional and global motion of the left ventricle (LV) of the heart. Early and fundamental signs of a CAD are believed to show in LV wall motion as abnormalities in one or several segments of the LV wall, where a segment may move "abnormally" or "non-uniformly". This abnormality can be defined as a "weak motion", known as *hypokinesia*, "no motion", known as *akinesia* or "out of sync", known as *dyskinesia*. The primary tool to detect, identify and quantify such regional wall motion abnormalities (RWMA) is the patient's echocardiogram (echo), which is notoriously difficult, and subjective and, therefore, highly operator dependent. Although RWMA is the first abnormality to set in with the onset of myocardial ischemia, preceding metabolic and electrocardiographic abnormalities, it is currently only used as a secondary diagnostic tool in patients with non-diagnostic ECG or when diagnosis is not evidenced (or shown/proven) by "standard" means, despite the fact that echo and particularly myocardial strain imaging provide an early diagnosis of an acute MI when RWMA is present. The reasons for not using echo as a first line diagnostic tool, for suspected MI patients, are largely interpretational. Echocardiographic interpretation, such as ultrasound scan, is highly operator-dependent as it depends upon the visual estimation of the left ventricle (LV) muscle-wall motion, its radial displacement and deformations. As a result, the final diagnosis suffers severely from the high inter-observer and intra-observer variability, making it prone to human errors and misjudgments.

To address these challenges, there is a need for an automated, robust and accurate tool that can assist cardiologists and echo-technicians understand and interpret echo more accurately, which may lead to saving lives. Despite the need, there are only few studies in the literature which proposed an automatic method for the LV wall motion estimation and abnormality detection from echo [3]. This is not surprising because first of all, there is no publicly available benchmark echo database with ground-truth labels. Second, capturing the global motion of an arbitrary shaped LV segment is difficult especially when the quality and/or spatial/temporal resolution of the echo is low. Finally, "motion estimation" in a video is known to be an *ill-posed* problem [3]-[5] even for natural videos with distinct objects/textures. This is true both for dense (pixel-based) or local (few pixels or block-based) motion estimation. Therefore, for a typical echo, which may be too noisy, estimating the true motion of the entire LV wall from such local (group of) pixels will be difficult and in some cases, impossible [6]. A recent study [7] has attempted to address this problem by two Machine Learning approaches. Both approaches have obtained low accuracies varying 57% to 85.4% and specificities varying 47% to 77.6% despite the fact that the echo videos were in relative high quality and pre-processed.

Due to these limitations and drawbacks, echocardiographic strain and strain-rate imaging (deformation imaging), instead, became the main focus of many studies as a non-invasive method

Serkan Kiranyaz (mkiranyaz@qu.edu.qa), Rayyan Ahmed, Rayaan Abouhasera, Junaid Malik (hafiz.malik@qu.edu.qa), and Ridha Hamila (hamila@qu.edu.qa) are with the Department of Engineering, Qatar University, Doha, Qatar. Tahir Hamid and Rashid Mazhar are with HMC, Doha, Qatar.

Aysen Degerli (aysen.degerli@tuni.fi), Morteza Zahibi (morteza.zahibi@tuni.fi), Moncef Gabbouj (moncef.gabbouj@tuni.fi) are with the Faculty of Information Technology and Communication Sciences, Tampere University, Tampere, Finland.

for the assessment of myocardial function. LV wall motion and wall deformation (strain) are different assessments. First of all, while the motion can be estimated and assessed by human experts in a subjective way, this is not possible for the strain because it represents the amount (and rate) of deformation of the LV wall such as longitudinal shortening (negative strain) and radial thickening (positive strain) during myocardial contraction. A human eye, no matter how trained and experienced the practitioner can be, cannot sense or measure this from an echo. On the other hand, strain and strain rate (SR) measurements are also derived from the myocardial velocities over the LV wall. The most common technique is called "Speckle Tracking" [6]-[23], which attempts to capture the motion by tracking "speckles" (natural acoustic markers) in the 2D ultrasonic image (echo). Speckles are the brightest patches and they usually are about 20 to 40 pixels. In prior studies, they are assumed to be "stable" from frame to frame. So, under the assumption of an accurate frame-by-frame tracking, the change of a speckle position gives its velocity and thus the LV wall motion is somewhat reflected by the motion of the speckles. They are chosen at the LV segment borders to produce the motion curves from which (negative) strain (i.e., shortening) and SR can be estimated. Therefore, the accuracy of the strain and SR estimation, too, solely depend on the accuracy of the motion estimation (and tracking) of each joint speckle during one or more cardiac cycles. Although the motion estimation of a local group of pixels is much easier than the estimation of the global motion of the entire LV wall, it still suffers from this *ill-posed* nature of the problem, e.g., due to small sensor movements, medium to very high noise levels, poor temporal and spatial resolution, and other sources of distortions [6]. For example, the minimum frame rate required for a reasonable speckle tracking is 60 fps. A higher frame rate is another challenge since it reduces the spatial resolution resulting in poor tracking. Moreover, robustness is a crucial issue since it is a well-known fact that different speckle tracking algorithms produce different results. In this study, we shall demonstrate that even when robust key-points are used instead of those speckles, robustness still remains the main problem. Eventually, the curves resulting from strain imaging are highly variable and their interpretation for diagnosis is subjective and experience-dependent [11]. Even under ideal conditions, many studies [11], [19]-[23] reported around 80-85% sensitivity rate for the detection of the infarcted segments. For instance, in an earlier study on 30 patients Leitmann *et al.* [19] found that 80.3% of the infarcted segments and 97.8% of normal segments were adequately recognized by speckle tracking based 2D-strain imaging. Even when the Doppler echo is used, the longitudinal Doppler strain data displayed 85% sensitivity and specificity for the detection of infarcted segments [20]. As a result, despite some promising and recently published studies [21]-[23], strain imaging based on speckle tracking is not ready yet for routine assessment of MI viability [11].

In this study, we propose a novel Computer Vision method, called *Active Polynomials* (APs) that can capture the global motion of the LV wall in a robust and accurate way. Our objective is to *mimic* an expert cardiologist who can analyze the echocardiogram records by visually searching for any RWMA for the early detection of an acute myocardial dysfunction. While the cardiologist can only perform this subjectively, APs can be used to capture and measure the *true* motion of the LV wall; therefore, it can identify and quantify the regional motion abnormalities. In order to accomplish this, APs are formed on the endocardial boundary of the LV wall and chamber. This boundary is the most promising salient feature of an echo where the maximum contrast usually occurs. In order to capture this boundary, we use the Active Contour (or snake) [24] with an artificial constraint embedded on the LV wall with the Ridge Polynomials (RPs). RPs will ensure a *converging* snake initiated in the LV chamber; however, due to the high noise level, the snake may still partially fail to capture the true endocardial boundary of the LV wall. This is why we shall then fit a $4^{th}$ order polynomial over the snake in order to obtain the APs that can cover the boundary of the entire LV wall in a smooth and continuous manner. Once this is repeated for all frames in the echo, then the global motion of the segments on the LV wall can be modelled by "motion activity" curves and their maximum displacement can be measured. While the APs can be used as an automatic tool to detect and identify objectively a possible segment motion abnormality (and hence to identify the infarcted segments causing a possible MI), it can also be used as an enhanced visualization platform over the raw echo to assist cardiologists or echo operators for a more accurate diagnosis and echo quality assessment. Finally, the proposed method is tested extensively over the first benchmark echo dataset, HMC-QU, solely created for this purpose by the physicians in Hamad Medical Corp. (HMC) Hospital and researchers in Qatar University (QU). HMC-QU encapsulates 160 echos, 89 of which are from acute MI patients and the rest from normal (non-MI) patients. HMC-QU is the first publicly available dataset and is the largest by far among the non-public ones used in prior works [6]-[23]. The 4-chamber view echos of the MI patients are labelled by a group of physicians in HMC Hospital, and the proposed method is evaluated based on these ground-truth segment labels of each echo.

The rest of the paper is organized as follows. Section II provides preliminary work on MI detection on echo and Active Contours. Section III presents the proposed LV wall motion estimation for MI detection and identification. The benchmark echo dataset, HMC-QU will first be introduced in Section IV. Then both quantitative and qualitative evaluations of the proposed approach over the HMC-QU dataset will be detailed and the MI detection performance will be analyzed together with a computational complexity analysis. Finally, Section V concludes the paper and suggests topics for future research.

## II. PRELIMINARIES

### A. MI Detection by Local Motion Estimation

MI is a major cause of death worldwide and gaining momentum especially during the last decade. In pathology, MI can be defined as myocardial cell death due to prolonged ischemia. When a coronary artery is blocked, the CAD shows the first signs of perfusion abnormalities due to the lack of oxygenated blood flow to the LV tissue within minutes from the occlusion of the coronary artery, leading to severe ischemia, which produces regional wall motion abnormalities (RWMAs). These RWMAs can be visualized by echo. This is the onset of a MI that can be even before the patient feels a chest pain or angina. That is why echo is an essential tool to detect the onset of myocardial ischemia and to identify the arteries with blockage.

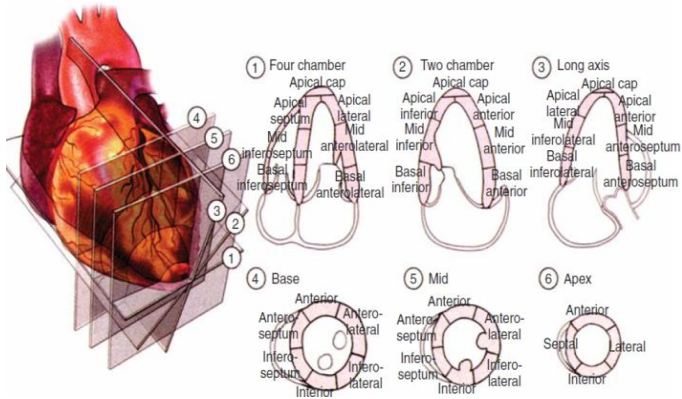

Figure 1: LV 17-segment model and respective views [18].

In an echo, there are different standardized LV segmentation models, such as 16-segment, 17-segment, and 18-segment models. The American Heart Association Writing Group on Myocardial Segmentation and Registration for Cardiac Imaging recommendation is to use the 17-segment model [18] shown in Figure 1. In this study, the proposed technique has been developed and extensively tested on the 4-chamber view; however, it can directly be used on the other views. As illustrated in Figure 2, in the 4-chamber view, the LV has 7 segments where 6 of them except the *apical cap* (segment-4) exhibit a uniform motion activity. Prior studies that attempt to compute the longitudinal strain by speckle-tracking echocardiography fix a speckle at each segment boundary and attempt to track it during one or few cardiac cycles. Due to the aforementioned limitations and drawbacks, even in ideal cases (i.e., low noise, high frame rate, and full contrast), the speckle tracking methods can achieve around 80-85% sensitivity and specificity levels.

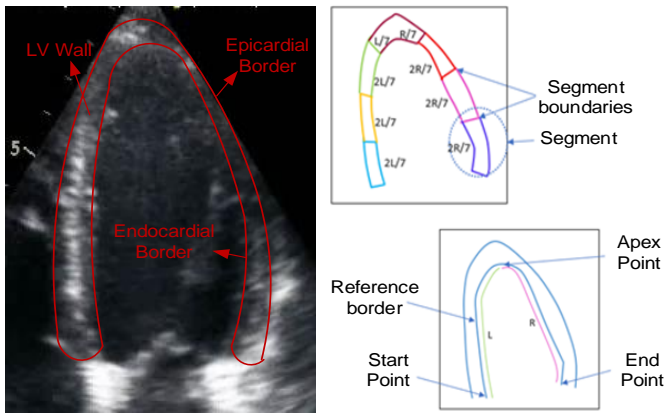

Figure 2: The LV wall and its borders (right) in the 4-chamber view. The segmentation of the LV wall and the start-end points (left).

In this study, we first investigated whether more stable and robust key-points can indeed cure the drawbacks of speckles. So, instead of a motion analysis based solely on a single speckle on each segment boundary along with its local region, we extract a large number of highly robust key-points on the LV wall by using the method called, "Speeded up Robust Feature" (SURF) [25], which belongs to the family of well-known key-point extractors in Computer Vision [25]-[28] including the first and perhaps the most popular one, "Scale Invariant Feature Transform" (SIFT). Juan and Gwun in [25] evaluated the performances of SIFT, PCA-SIFT and SURF methods for scale, rotation, and affine transforms as well as for blur and illumination changes. This study has shown that SIFT performed slightly superior in most experiments but with the slowest speed (highest computational complexity). Some other experiments have shown that SURF was the fastest and the most stable [26]. Obviously, both SIFT and SURF points show a superior robustness over the naïve speckles with the sole feature of "high brightness" which may change abruptly due to noise, sensor disturbance or other possible factors.

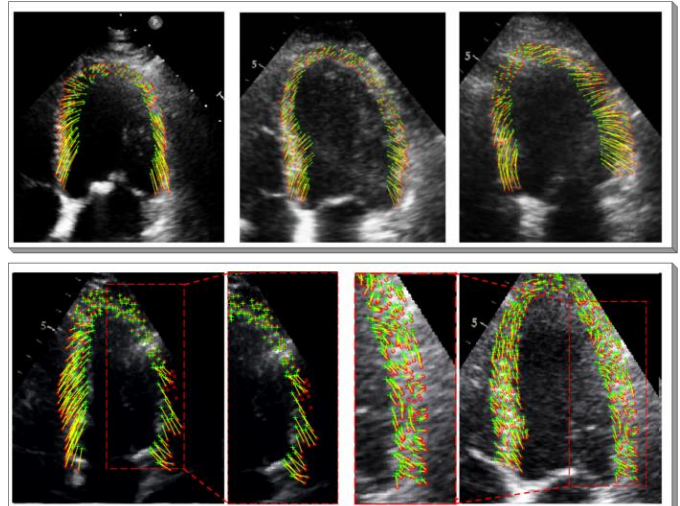

Figure 3: Accurate (top) vs. erroneous (bottom) tracking of the SURF points on the LV wall.

Our aim is to investigate whether a large number of key-points can indeed be used to capture the global motion of the LV wall. Accordingly, we can also find out whether they can be used to compute the strain in a robust and accurate manner. For the former, the results have shown that especially when the noise level is high, even the majority of the robust SURF points may lead to erroneous tracking as shown in Figure 3. While SURF points are coherent and able to capture the global motion for those echos on top of the figure, the zoomed sections of the echos on the bottom clearly show that the majority of the SURF points, despite their robust and stable nature, could not be tracked due to the high level of noise appearing on the next frame. Obviously for the latter aim, tracking of SURF points on the boundaries of the segments may fail too and this will result in erroneous strain computations, which in turn yield misdiagnosis of the heart status. This shows that such "bottom-up" approaches to capture the true motion of the LV wall using the local key-points may neither be robust nor reliable for MI detection. This clearly indicates that the global motion should instead be captured in a "top-down" fashion. The two possible solutions for this approach are the (accurate) segmentation of the LV wall or extraction of the entire endocardial boundary at each frame of the echo. There are few recent attempts for the former using recent Deep Learning paradigms [29]-[32]; however, they cannot still guarantee an accurate segmentation especially when the echo quality is poor. In this study, we shall focus on the latter, the extraction of the endocardial boundary, which can indeed be performed with a high accuracy using the proposed approach. The starting point for this is the Active Contours [24], which will be reviewed next.

## B. Active Contours

An active contour (or snake) is an elastic 2D spline whose contour is guided by internal (smoothness and curvature) and external (image gradients and edges) constraints. The problem is transferred to the minimization of a joint (total) energy, $E_T$, that can be expressed as follows:

$$E_T = E_I + \gamma E_X = \int_s (E_I(v(s)) + \gamma E_X(v(s))ds. \qquad (1)$$

where $E_I$ is internal and $E_X$ is external energy terms that define the respective constraints, $\gamma$ is the regularization coefficient. In this study, we used a more recent and improved version of snake [33] the details of which are covered in Supplementary (A).

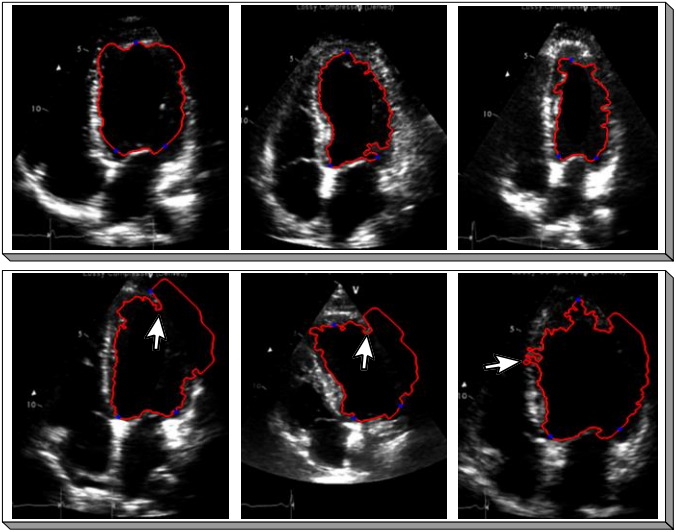

Figure 4: Snake method for LV segmentation on 6 echos. Reasonable (top) vs. erroneous (bottom) results.

Snake method has been directly used for LV segmentation in a recent study [34]; however, it has only been tested on the frames of a single echo. Although the result was satisfactory, obviously such a limited evaluation is not sufficient. Especially when the quality of echo degrades, the snake may fail to converge to the true boundary of the LV wall. Typical examples can be seen in Figure 4 (bottom) where the snake not only failed to converge to the true boundary due to lack of contrast, it also presents severe noise sensitivity on the boundaries as shown by the white arrows. This basically demonstrates the fact that the "snake-only" approach cannot exhibit the required robustness and accuracy to capture the LV endocardial boundary along with its global motion. In the next section, we shall detail how the proposed method addresses effectively this drawback by using the proposed approach with Active Polynomials (APs).

## III. METHODOLOGY

The proposed method consists of two consecutive phases, as illustrated in Figure 5. The first phase is the LV wall extraction. In the second phase, using the APs formed on the LV wall, 7 segments are extracted for MI detection, identification and for further enhanced visualization capabilities to assist cardiologists perform their diagnosis. In the following sections, we shall detail each phase.

## A. LV Wall Extraction

As illustrated in Figure 5, the formation of Active Polynomials (APs) is performed in three stages over each frame of the echo. The first stage is the formation of an artificial wall, Ridge Polynomials (RPs), to prevent the divergence of the snake. The second stage is the formation of the snake within the LV. Finally, the third stage is the composition of the APs over the snake. In the following subsections, we shall detail each stage.

### 1) Ridge Polynomials

The proposed approach to capture the endocardial boundary is designed to address the two drawbacks of the active contours on echos. The first and the foremost problem is the partial divergence of the snake when the contrast is poor, e.g., see the three echos in Figure 4 (bottom) where the snake fails to converge to the boundaries of segments 4, 5 and 6. To prevent this, we artificially enhance the contrast by building a white wall on top of the brightest section (i.e., the ridge) of the LV wall. In those problematic echos in the figure, the ridge can even be invisible to the naked eye due to the lack of contrast; however, it still exists with low brightness values while still having the maximum intensity in a local neighborhood. So, the idea is to build a sufficiently thick wall (e.g. 6 pixels) by incrementally increasing the intensity value (e.g. 200 to 255 for 8-bit image representation) on top of the connected series of brightest pixels (i.e. the ridge). To obtain the anchor points on the ridge, we use the start and end points as illustrated in Figure 2 along with the topmost point of the *apical cap*. As shown in Figure 6 considering that the LV boundary is divided into left and right parts equally from the apex point, two lines are fitted from the top to the start and end points. The ridge (maximum intensity) points are detected as moving the 14 equally distanced anchor points towards the boundary. New set of anchor points are defined by stretching the anchor points horizontally to obtain a pair of left and right anchor points shown in Figure 6. Finally, over the left and right anchor points, two $4^{th}$ order ridge polynomials (RPs) are initially fit which will constitute the borders of the search region, i.e., the "Region of Interest" (RoI) inside of which the actual ridge points will be searched. Once the ridge points are detected, the $4^{th}$ order ridge polynomial is fit to these ridge points using the regularized Least-Square (LS) optimization. The details of fitting an $n^{th}$ order polynomial over $m > n+1$ points using regularized LS method is given in Supplementary (B). The RPs are then used to create the artificial wall (barrier) so that the snake is guaranteed to converge to the LV wall.

### 2) The Constrained Snake

As illustrated in Figure 7 (left), the snake is initialized as a mini-form of the RPs within the LV wall which is encapsulated by the actual RPs. After 300 iterations, the snake converges to the true endocardial boundary of the LV wall in Figure 7 (right). In this particular echo, without the artificial wall made by the two RPs, the snake would have diverged on the right side of the LV wall due to lack of contrast of the echo shown in this figure. In this echo and on several others where the snake is divergent, the artificial wall solved this problem. However, the second drawback of the snake approach, the high noise sensitivity, is still evident. The snake may fail to converge to the true boundary due to the noisy speckles within the blood chamber. Furthermore, excess noise level usually makes the snake unnecessarily detailed at the boundaries as shown in Figure 4 (white arrows at the bottom). In order to address these drawbacks, the formation of the proposed Active Polynomials (APs) will be detailed in the next sub-section.

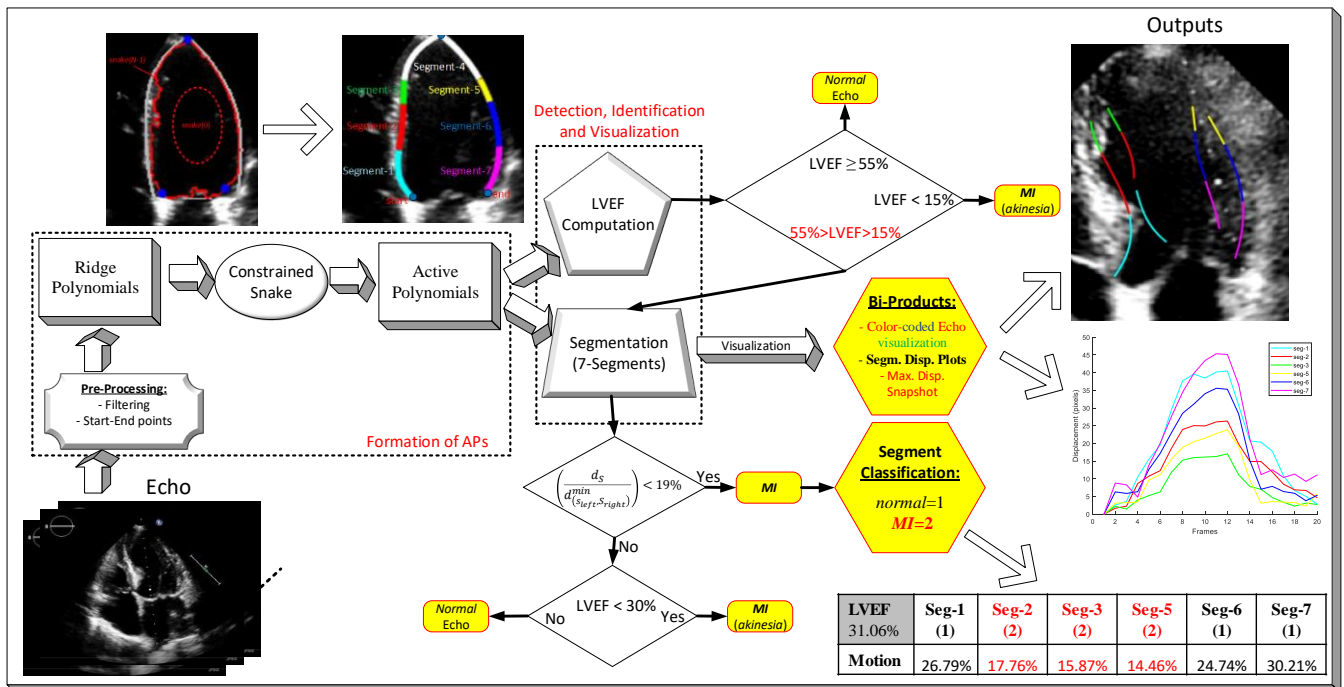

Figure 5: The overview of the proposed method for MI detection and identification. The yellow shaded blocks generate the outputs on the right-most side.

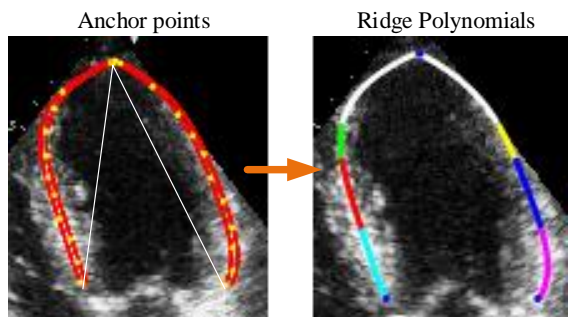

Figure 6: (left) The anchor points (yellow) on LV wall and (right) final ridge polynomial (RP).

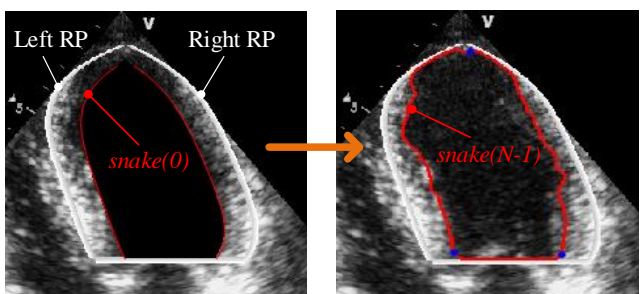

Figure 7: Initialization of the snake inside the two RPs (t=0) and the final snake obtained after N = 300 iterations.

### 3) Active Polynomials

Although some sections of the snake suffer from occasional "over-fitting" problem possibly due to the excess noise, even such a problematic snake can still serve as the initial "reference" to capture the endocardial boundary with smooth polynomials. For this purpose, a pair of 4th order polynomials, for the left and right side of the LV wall are fit to the points of the snake using the regularized LS. For the left one, the 9 equally spaced snake points between the *start* to *apex*, and for the right one, between *apex* and the *end* are used. Since each polynomial will assume a smooth shape of the active contour, we call them Active Polynomials (APs). As shown in Figure 8, the two parts of the snake (purple and yellow) are used to compose a pair of APs and finally, they are used to create 7 segments (segments 1 to 7 counter-clockwise) of the 4-chamber view echo.

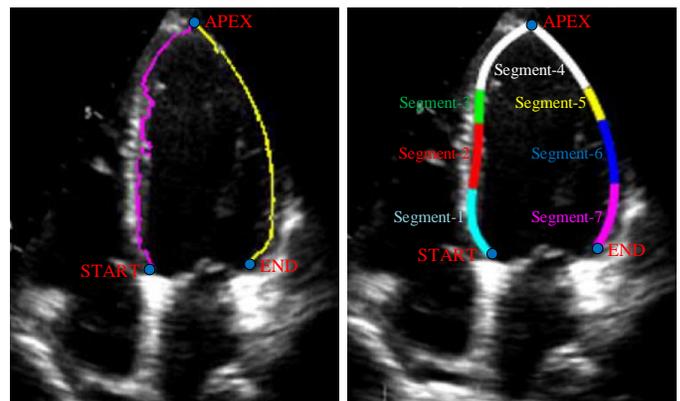

Figure 8: Snake points (left) are used to create a pair of APs (right) that are used to create 7 segments of the 4-chamber view.

### B. MI Detection, Identification and Visualization

This is the second block in Figure 5 which uses the output of the first block, APs, to perform the global motion analysis. APs are divided into 7 segments as shown in Figure 8 (right) and their movement (displacement) is monitored. Once the global motion of each segment is captured by simply evaluating the "rate of displacement", we can mimic a typical cardiologist's diagnosis of a motion anomaly by detecting which segment or segments are showing signs of abnormal (non-uniformity or lack of) motion activity. However, before going into motion analysis, the LV Ejection-Fraction (LVEF) ratio is first computed as follows:

$$LVEF = \frac{EDV - ESV}{EDV} \approx 1 - \frac{A_{min}}{A_{max}} \qquad (2)$$

where $EDV$ and $ESV$ are the end-diastolic and end-systolic volumes, respectively. In a 2D echo, one can estimate them by computing $A_{min}$ and $A_{max}$, which are the minimum and maximum area of the LV chamber, respectively. They are proportional to the total number of pixels encapsulated by the snake or by the two APs. The recommendation for *LVEF* to indicate a "reference" (normal) and "severely abnormal" LV activities for both men and women, are LVEF ≥ 55% and LVEF < 30%, respectively [35]. Following this recommendation, the proposed motion analysis will no longer be performed when LVEF ≥ 55% and the echo can directly be classified as *normal*. However, for the lower limit, we empirically use a more conservative threshold, LVEF ≤ 15, which is obviously a sensitive marker of myocardial dysfunction and a clear sign of MI. For this severe case of myocardial dysfunction, the echo with all the segments can directly be classified as MI. Therefore, motion analysis will *only* be performed when 55% > LVEF > 15% and the outcome of the motion analysis will determine whether the echo is normal or MI. In this case, if there is *at least* one myocardial segment with abnormal (*hypokinesia* or *akinesia*) motion activity then MI is detected and the corresponding arteries with blockage can be identified.

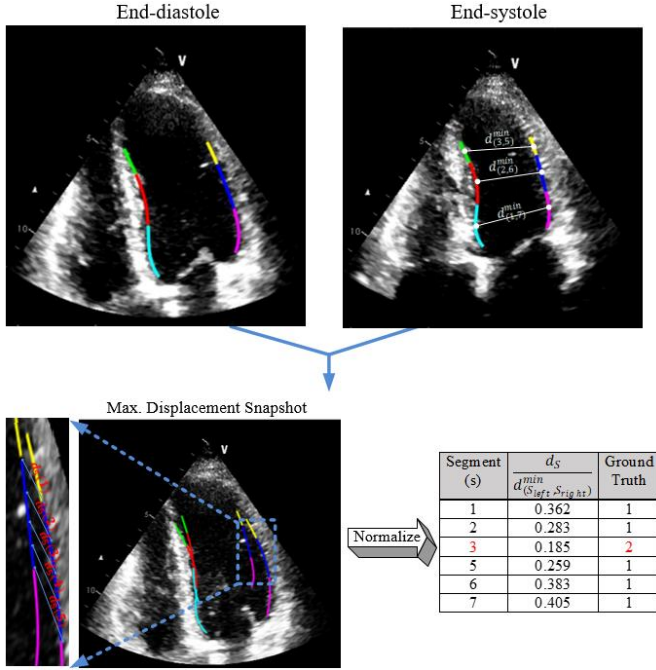

Figure 9: Computation of the normalized maximum displacement of the 6 segments of the 4-chamber view echo with the ground-truth labels (normal = 1, infarcted = 2)

When a cardiologist visually evaluates the motion activity of a 4-chamber view echo, the infarcted segments that show a "reduced" motion (or almost no motion at all) compared to other segments are identified either as *hypokinesia* or *akinesia*. The motion assessment is obviously independent from the resolution of the echo. The study over the segments labeled as *abnormal* by the cardiologists in HMC-QU database has indicated that those segments that move less than 20% of the minimum interval to the corresponding segment on the other side of the chamber are diagnosed as MI. Since the minimum interval between corresponding segments is resolution dependent, the ratio of maximum displacement to this minimum interval will therefore allow us to mimic the cardiologist's evaluation in a quantitative way.

Let $d_S$ be the maximum displacement of the segment, $S$ where $S \in \{1,2,3,5,6,7\}$. Let $d_{(S_{left},S_{right})}^{min}$ be the minimum interval during a cardiac cycle of an echo of the corresponding segments, $S_{left}$ and $S_{right}$ where $S_{left} = 1,2$ or $3$ and $S_{right} = 5,6$ or $7$, respectively. Since the maximum motion (of a segment), $d_S$, is proportional to the (maximum) displacement occurred from *end-diastole* to *end-systole*, the proposed method computes $d_S$ as the maximum displacement of each segment and normalizes them by $d_{(S_{left},S_{right})}^{min}$. Both measurements are one-pixel accurate and the ratio is then compared with an empirical threshold (e.g., 19%) that is selected just below 20%.

Figure 9 illustrates a sample echo from a MI patient where cardiologists labeled segment 3 as *akinetic*, and the rest as *normal*. The two APs extracted for the frames corresponding to *end-diastole* and *end-systole* are shown in the figure. Knowing the *end-diastole* as the first frame, one can easily find the frame of the *end-systole* by simply searching for the maximum overall segment displacement. However, in this study instead of considering the segments in the *end-systole* frame, we search for the maximum displacement of each individual segment, which may not necessarily come from the *end-systole* frame. Experiments show that most of the segment-wise maximum displacement indeed occurs at the *end-systole* frame; however, occasionally it may also occur at the frames within a close vicinity (e.g., ±1-2 frames). It is straightforward to notice the "reduced" motion from the gap between $\left(\frac{d_S}{d_{(S_{left},S_{right})}^{min}}\right)$, ratios of segment 2 and the rest where the latter group has ratios above 19%. Therefore, the results are in full agreement with the ground-truth labels made by the cardiologists.

In order to compute the maximum displacement of a segment, at first, uniformly sampled $n_S$ points are taken over the segment and then the segment displacement can be approximated by averaging the point-wise distances, as expressed below for the 2$^{nd}$ segment ($S$=2) shown in the figure (e.g., for $n_S = 5$).

$$d_S = \frac{1}{n_S} \sum_{i=1}^{n_S} d_S(i) \qquad (3)$$

where $d_S(i)$ is the $i^{th}$ point's maximum displacement of the segment $S$=2 as shown in the figure. There are several options to compute individual point-wise distances, $d_S(i)$. When they visually assess echo, cardiologists consider the motion in both x and y directions. Therefore, we can use $L_1$, $L_2$ or $L_\infty$ norms to compute $d_S$ which are expressed as below.

$$d_{L_1} = |x_2 - x_1| + |y_2 - y_1|$$

$$d_{L_2} = \sqrt{(x_2 - x_1)^2 + (y_2 - y_1)^2} \qquad (4)$$

$$d_{L_\infty} = max(|x_2 - x_1|, |y_2 - y_1|)$$

where $(x_1, y_1)$ and $(x_2, y_2)$ are the x- and y-coordinates of the corresponding $i^{th}$ points on the segment at the *end-diastole* and *end-systole* frames, respectively. A closer look will reveal the fact that $d_{L_\infty}$ does not exactly mimic the aforementioned way the cardiologist assess motion since $d_{L\infty}$ only reflects the distance in either x- or y-direction. Between the remaining norms, we use the $L_2$ norm since it is the natural distance metric for the human perception.

An alternative way to assess the segment motion is the segment displacement plots as shown in Figure 10, which show the displacement of each segment during one cardiac cycle. This plot is in fact more informative than the maximum motion ratios given in Figure 9 because the instantaneous and average motion of each segment can also be computed besides their maximum displacements. However, we still perform our motion analysis based on the maximum displacement due to the simple fact that the derivative operator is noise sensitive and the displacement curves will inevitably bear certain level of measurement noise.

Finally, the proposed method presents several enhanced visualization options that will significantly assist medical experts perform their diagnosis. For instance, the color-coded segments formed over the two APs provide a cardiologist with a better motion estimation than the one from the raw (gray-scale) echo since the cardiologist can now see, distinguish and assess each individual segment displacement and (instantaneous) motion in a visually enhanced manner. Another bi-product of the proposed method is the maximum displacement snapshot as shown in Figure 9, which allows cardiologists to visualize the displacement anomaly.

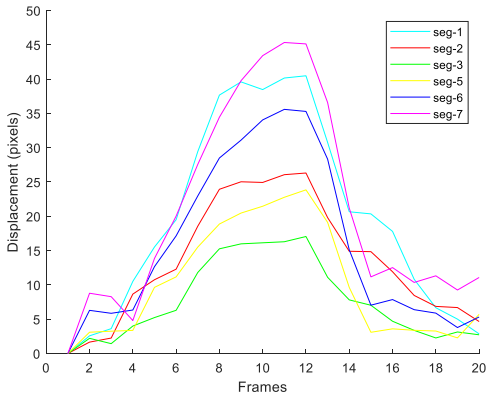

Figure 10: Segment displacement plots for the echo shown in Figure 9 with respect to the first (*end-diastole*) frame.

## IV. EXPERIMENTAL RESULTS

### A. HMC-QU Benchmark Dataset

HMC-QU benchmark echo dataset has been created by collaboration between Qatar University (QU) and Hamad Medical Corporation (HMC) Hospital. HMC-QU contains 160 4-chamber view echo recordings obtained at the HMC hospital between 2018 and 2019. These cases are from over 10000 echos performed in a year including more than 800 cases admitted with acute ST elevation Myocardial infarction. The echos included in our assessment belonging to the 89 MI patients (all first time and acute MI) and the rest are normal. There are 13 women and 76 men in the MI patient group. All MI echos were obtained from patients who were admitted with a diagnosis of acute MI with evidence obtained from ECG, cardiac enzymes and who underwent coronary angiogram/angioplasty to treat the MI. These patients had echos obtained within 24 hours of admission or in some cases before they underwent coronary angioplasty. All "Normal" echos were defined, as the echos of the patients not admitted for MI (acute or previous) but acquired for other reasons including health check and investigation of murmurs. All 4-chamber view echos have been labelled segment-wise by cardiologists in HMC hospital. The 6 segments of 4-chamber view of each echo is labelled as: *normal*=1, *hypokinetic*=2 and *akinetic*=3. Echos are acquired by devices from different vendors, such as Phillips Ultrasound machines and GE Vivid (GE-Healthcare-USA) Ultrasound Machine. The temporal resolution (frame rate per second) of the echos is 25 fps. The spatial resolution also varies from 422×636 to 768×1024 pixels. The duration of each echo taken for analysis is one cardiac cycle.

### B. Results

In this study, each echo is categorized as *normal* or *MI* while each segment in a 4-chamber view echo is categorized as *normal* (1) or *infarcted* (either *hypokinesia* = 2 or *akinesia* = 3). As illustrated in Figure 5 the motion analysis is only performed if 15% < LVEF < 55%. If LVEF ≥ 55%, all segments are assumed to be *normal* (1) and if LVEF ≤ 15% all segments are assumed to be *akinetic*. Otherwise, the motion analysis will determine whether the echo is *normal* or *MI*. If there is at least one infarcted segment with abnormal motion activity, then the echo is assumed to be *MI*; otherwise, *normal*. A segment is assumed to be *infarcted* if its motion ratio is below 19%. The thresholds, 55% and 30% for LVEF are recommended in [35]. Once all echos in the dataset together with their segments are categorized by the proposed algorithm, then the confusion matrices (CM) are formed by evaluating the assigned categories with respect to the ground-truth labels. This enables us to compute the following standard performance metrics for MI detection and infarcted segment identification performances: classification accuracy (*Acc*), sensitivity (*Sen*), specificity (*Spe*), and positive predictivity (*Ppr*). CM elements are the hit/miss counters such as true positive (*TP*), true negative (*TN*), false positive (*FP*), and false negative (*FN*). The following standard performance metrics can now be expressed using them: *accuracy* is the ratio of the number of correctly detected echos (or segments) to the total number of echos (segments); *sensitivity* (or *Recall*) is the rate of correctly detected MI echos among all MI echos in the dataset; *specificity* is the *sensitivity* of the normal echos; and, *positive predictivity* (or *Precision*) is the rate of correctly detected MI echos in all the echos detected as MI. Finally, the *false alarm rate* (*FAR*) can be defined as: $FAR = 1 - Spe$.

Figure 11 shows the *end-diastole*, middle and *end-systole* frames of a *normal* and *MI* echos where the 7 segments are color-coded over the two APs along with their maximum displacement snapshots. The relative motion (displacement) of the segments 1, 2, 3 and 5 makes it straightforward to detect the motion abnormality on the MI echo while apparently all segments of the *normal* echo move in a uniform manner. Despite the fact that the quality is quite poor with a significant noise in the interior (blood) chamber and the temporal resolution is low (25 fps) in both echos, the proposed method successfully captures the global motion of the LV wall.

The MI detection and identification (detection of the segments with abnormal motion activity) performances are presented in Table 1 (per segment), Table 2 (over all segments) and Table 3 (over all echos), respectively. Several important observations can be made based on these results. First of all, all results are based on the selected motion ratio threshold, 19%. No optimization or fine-tuning was performed on this threshold for maximizing certain criteria. The high accuracies achieved for detecting infarcted segments and echos approve the validity of this threshold; however, there is still room for improvement. The most crucial performance criterion is of course *sensitivity* (*Recall*) for infarcted segments and especially MI echos. Especially the latter, *Sen(MI) > 91%*, indicates an elegant performance level considering the low temporal resolution and the poor quality of many echos in the dataset. The secondary objective is to minimize the false alarms (or equivalently to maximize the *specificity*). This is a misdiagnosis case for MI, which can be corrected by the cardiologists when the proposed method causes a false alarm. The *FAR* was rather low ( i.e., < 9%) for detecting segment motion abnormality; however, the results in Table 3 show a rather high *FAR*. The main reason is that mis-diagnosing a normal segment as infarcted suffices to misclassify the echo as MI. On the other hand, the fixed threshold used for detection, 19%, may yield such misdiagnosis because the proposed method has a certain sensitivity for capturing the global motion, i.e., in the vicinity of ±2 pixels and, therefore, a slight variation from the actual displacement may cause such a misclassification even if it is as low as 1-2 pixels, e.g. assume that for a normal segment, $d_{(s_{left}, s_{right})}^{min} = 100 \ pixels$ and $d_S$ is measured as 18 pixels with +2 pixels bias. This will cause a false alarm since $\left( \frac{d_S}{d_{(s_{left}, s_{right})}^{min}} \right) = 18\% < 19\%$.

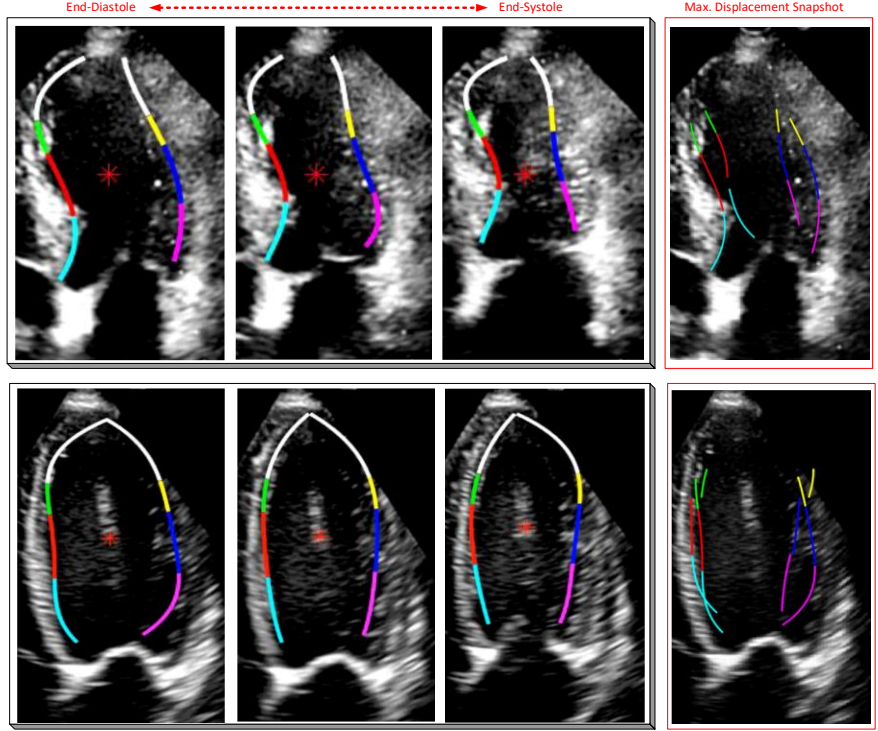

Figure 11: End-diastole, middle and end-systole frames of a normal (top) and MI (bottom) echos. Their maximum displacement snapshots are shown on the right.

Table 1: The performance of the proposed method (per-segment) on detecting the infarcted segments.

| Segment | Sensitivity | Specificity | False alarm | Precision | F1-score | Accuracy |
|---------|-------------|-------------|-------------|-----------|----------|----------|
| 1 | 0.8966 | 0.9389 | 0.061 | 0.7647 | 0.8254 | 0.9313 |
| 2 | 0.7843 | 0.8807 | 0.1193 | 0.7547 | 0.7692 | 0.85 |
| 3 | 0.8676 | 0.8587 | 0.1413 | 0.8194 | 0.8429 | 0.8625 |
| 5 | 0.8545 | 0.9362 | 0.0638 | 0.8868 | 0.8704 | 0.906 |
| 6 | 0.6875 | 0.9187 | 0.0813 | 0.6875 | 0.6875 | 0.871 |
| 7 | 0.7 | 0.9348 | 0.0652 | 0.6087 | 0.6512 | 0.9051 |

Table 2: The performance of the proposed method on detecting the infarcted segments.

| Sensitivity | Specificity | False alarm | Precision | F1-score | Accuracy |
|-------------|-------------|-------------|-----------|----------|----------|
| 0.8157 | 0.9141 | 0.0859 | 0.7790 | 0.7969 | 0.8875 |

Table 3: MI detection performance of the proposed method over the HMC-QU benchmark dataset with 160 echos.

| Sensitivity | Specificity | False alarm | Precision | F1-score | Accuracy |
|-------------|-------------|-------------|-----------|----------|----------|
| 0.9101 | 0.7606 | 0.2394 | 0.8265 | 0.8663 | 0.8438 |

Table 4: The performance of the proposed method (per-segment) on detecting the infarcted segments over echos with a reasonable quality.

| Segment | Sensitivity | Specificity | False alarm | Precision | F1-score | Accuracy |
|---------|-------------|-------------|-------------|-----------|----------|----------|
| 1 | 0.9286 | 0.9469 | 0.0531 | 0.8125 | 0.8667 | 0.9433 |
| 2 | 0.8125 | 0.8817 | 0.1183 | 0.7800 | 0.7959 | 0.8582 |
| 3 | 0.9048 | 0.8974 | 0.1026 | 0.8769 | 0.8906 | 0.9007 |
| 5 | 0.9020 | 0.9444 | 0.0556 | 0.9020 | 0.9020 | 0.9291 |
| 6 | 0.7586 | 0.9464 | 0.0536 | 0.7857 | 0.7719 | 0.9078 |
| 7 | 0.7647 | 0.9435 | 0.0565 | 0.6500 | 0.7027 | 0.9220 |

Table 5: The performance of the proposed method on detecting the infarcted segments over echos with a reasonable quality.

| Sensitivity | Specificity | False alarm | Precision | F1-score | Accuracy |
|-------------|-------------|-------------|-----------|----------|----------|
| 0.8602 | 0.9295 | 0.0705 | 0.8252 | 0.8423 | 0.9102 |

Table 6: MI detection performance of the proposed method over echos with a reasonable quality in the HMC-QU benchmark dataset.

| Sensitivity | Specificity | False alarm | Precision | F1-score | Accuracy |
|-------------|-------------|-------------|-----------|----------|----------|
| 0.9512 | 0.8136 | 0.1864 | 0.8764 | 0.9123 | 0.8936 |

The main reason for some of the false alarms encountered was the extremely poor image quality of some of the echos in the dataset (19 out of 160), which degrades significantly the actual performance level of the proposed method. In such cases, even an expert Cardiologist may not perform an accurate diagnosis on these echos. The reason we have included them in the dataset is to accomplish a realistic case and show the main source of misdiagnosis. When those 19 echos with such poor quality are excluded from the evaluation, the actual performance of the proposed method is presented in Table 4, Table 5 and Table 6. Finally, the poor temporal resolution is also the common drawback among all the echos and as discussed earlier, this, alone, is sufficient to render out the usage of any method based on Speckle Tracking or local motion estimation. The results presented in the above tables indicate that the proposed approach is quite robust against this drawback and certain level of quality degradations.

### C. Computational Complexity Analysis

Due to the unoptimized and sequential execution of the proposed method, its computational complexity is the sum of the individual computational complexities of the individual blocks illustrated in Figure 5. Please refer to Supplementary (C) for the computational complexity of each block along with the overall computational times in an unoptimized implementation.

### V. CONCLUSIONS AND FUTURE WORK

In this study, we propose a global method for finding the true motion of the LV by Active Polynomials which are formed at the endocardial boundary of the LV wall for each frame of an echo. Since the proposed method does not depend on local motion estimation and tracking, it is largely immune to the well-known ill-posed behavior and noise sensitivity of the 2D motion estimation. The proposed method is designed to "mimic" an expert cardiologist to capture the global motion in a similar manner so as to assess the regional motion with respect to the motion uniformity. The global extraction of the true motion of the LV wall enables to detect and identify the regional wall motion abnormalities (RWMA) which in turn can diagnose a myocardial infraction (MI) in an objective way. Since echo is the primary tool that can indicate the onset of a Myocardial Ischemia long before the ECG, this will help the detection of a MI at the earliest possible stages -practically as soon as the echo is acquired. Moreover, the proposed method voids the subjectivity and operator dependability of the echo interpretation and assessment since it can quantify the true measures of LV wall motion, (maximum) displacement and LVEF. Besides a standalone diagnostic tool, the proposed method can also offer several assistive bi-products such as enhanced visualization capabilities by color-coded APs over the raw echo, a snapshot of the maximum segment displacements to localize the motion abnormalities, segment displacement plots that can provide a deeper motion analysis and even a quality assessment tool for the echo acquisition. The last feature is especially important since an echo technician can fine-tune the echo probe manually until the proposed method can successfully compose the APs over the LV wall boundary. Especially, when the diagnosis of the proposed method and the (group of) cardiologist(s) differs, it can also be used as a "verification" tool that can give a second chance to the cardiologists to re-assess the cases or segments with mismatching diagnosis. Finally, when a range is used instead of a fixed threshold for classifying the segments, the proposed method can draw the attention for the *Unsure* cases where the segment motion is not definitive. Cardiologists can handle such difficult cases and make the final decision perhaps by assessing other echo views.

A crucial objective of this study is to achieve an utmost robustness against the high noise level in echos with a poor temporal resolution. Experiments over such echos show that this objective has been well-accomplished. This is particularly important because it is infeasible to analyze such echos with the current state-of-the-art MI detection methods such as "Speckle Tracking" or any other method based on local motion estimation. This has been verified in this study by a case study using the SURF key-points for tracking. Finally, for a proper performance evaluation, the HMC-QU dataset contains ground-truth labels. This is not only the first benchmark dataset that will be publicly available for the research community, it is also the largest collection ever compiled which consists of normal echos and echos of both male and female acute MI patients with different ages. An extensive set of experiments over this benchmark dataset demonstrates that the proposed method achieves an elegant sensitivity and precision for detecting MI and diagnosing the RWMA. For the latter, specificity is quite high yielding a low false-alarm rate. However, for the MI detection, the proposed method yields a significant false alarm rate; although not as severe as the false negatives, we aim to reduce false alarms with the joint analysis of the other views. Moreover, we aim to improve the speed of the proposed method by parallel computing paradigms and optimized implementation to achieve real-time analysis of the acquired echo. We expect to achieve an even better detection and identification performance with a proper analysis on the motion curves using a Machine Learning approach. These will be the topics of our future work.

## SUPPLEMENTARY MATERIAL

### A. Active Contours

The snake, $s$, at time, $t$, consists of finite number of vertices, $v(s)$ as shown in Figure 12, and each vertex will move (slowly) to another state (location) in order to minimize the total energy, $E_T$.

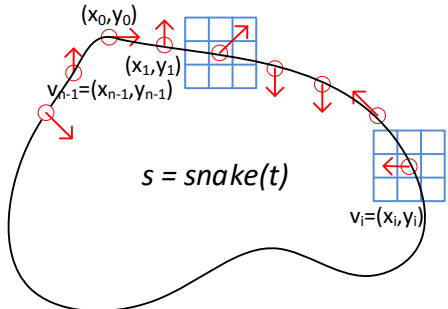

Figure 12: A snake with $n$ vertices at iteration, $t$. The next state of each vertex is shown with a red arrow.

The internal energy measures the bending (elasticity and stiffness) cost that can be expressed in terms of vertices, $v(s)$ over the snake, $s$, as follows:

$$E_I\big(v(s)\big) = \alpha \left|\frac{dv}{ds}\right|^2 + \beta \left|\frac{d^2v}{d^2s}\right|^2 \qquad (5)$$

Where for simplicity $\alpha$ and $\beta$ can be assumed to be constant parameters and the 1$^{st}$ and 2$^{nd}$ derivative can be approximated as,

$$\frac{dv}{ds} \cong v_{i+1} - v_i \ \ and \ \ \frac{d^2v}{d^2s} = \frac{d}{ds}\left(\frac{dv}{ds}\right)$$

$$\cong (v_{i+1} - v_i) - (v_i - v_{i-1}) \qquad (6)$$

$$\cong v_{i+1} - 2v_i + v_{i-1}$$

Note that, the term, $\frac{dv}{ds}$, forces for a "smooth" spline whereas the term, $\frac{d^2v}{d^2s}$, forces for the minimal curvature (no corners and discontinuous edges). Setting, $\beta = 0$ will, therefore, allow snake to converge to corners/edges. The reason that this energy term is called "internal" is because both derivatives depends only on the relative position of the consecutive vertices; i.e., there is nothing about the image intensity or gradient values. The second energy term, $E_X$, is designed to force the snake to converge towards certain image attributes such as "light/dark" objects or "object boundaries". Since in this study, we aim to capture the LV wall endocardial boundary, we shall set this term to the negative gradient of the image to minimize the external energy (to maximize the likelihood of the edge pixels to constitute the snake). This yields,

$$E_X\big(v(s)\big) = |G_x(v_i)|^2 + |G_y(v_i)|^2 \qquad (7)$$

As a result, in 2D image grid of an echo frame, Eq. (1) can be expressed as,

$$E_T = E_I + \gamma E_X =$$

$$\alpha \sum_{i=0}^{n-1} v_{i+1} - v_i + \beta \sum_{i=0}^{n-1} v_{i+1} - 2v_i - v_{i-1} + \qquad (8)$$

$$\gamma \sum_{i=0}^{n-1} |G_x(v_i)|^2 + |G_y(v_i)|^2.$$

The objective is to find the snake, $v(s)$, which yields the minimum total energy, i.e.,

$$v^*(s) = \arg\min_{v(s)} E_T\big(v(s)\big) \qquad (9)$$

Using the Calculus of Variations, it can be shown that the states of the minimum total energy will correspond to the zero crossings of the Euler-Lagrange equation that can be expressed as,

$$\alpha \frac{d^2v}{d^2s} + \beta \frac{d^4v}{d^4s} - \gamma \nabla E_X\big(v(s)\big) = 0 \qquad (10)$$

In order to create an iterative solution starting from an initial (estimate) of the snake, $v(s,0)$, we can parametrize Euler-Lagrange equation with time $v(s) \rightarrow v(s,t)$, the motion of convergence to a (local) optimum can be expressed,

$$\frac{dv(s,t)}{dt} = \alpha \frac{d^2v(s,t)}{d^2s} + \beta \frac{d^4v(s,t)}{d^4s} - \gamma \nabla E_X\big(v(s,t)\big)$$

$$\therefore \frac{dv(s,t)}{dt} - \alpha \frac{d^2v(s,t)}{d^2s} - \beta \frac{d^4v(s,t)}{d^4s} \qquad (11)$$

$$+ \gamma \nabla E_X\big(v(s,t)\big) = 0$$

Eq. (11) can then be solved by the discrete-time approximation and Linear Algebra which is used in this study. Alternatively, there are various methods to find the optimal state of the snake such as Gradient Descent, greedy search, evolutionary search, Dynamic Programming, etc.

In this study, we used a recent variation of snake method proposed in Chan-Vese, [33]. In this variant, the problem defined by [33] is the minimization of an energy-based segmentation. Consider a bounded open subset $\Omega$ of $\mathbb{R}^2$, the image $u_0$: $\overline{\Omega} \rightarrow \mathbb{R}$, the evolving curve $C$ in $\Omega$, as the boundary of an open subset $\omega$ of $\Omega$. Then, the $inside(C)$ represents the region $\omega$, and $outside(C)$ is the region $\Omega\backslash\overline{\omega}$. The boundary of the image $u_0$ is denoted as $C_0$. Then, the fitting term, $F$ is defined as in Eq.(12) where $C$ the snake curve, and the variables is $c_1, c_2$ are the averages of $u_0$ inside $C$ and outside $C$, respectively.

$$F_1(C) + F_2(C) = \int_{inside(C)} |u_0(x,y) - c_1|^2 \, dxdy$$

$$+ \int_{outside(C)} |u_0(x,y) - c_2|^2 \, dxdy \quad (12)$$

In this case, the optimal curve is the boundary of the image $C_0$ which is also the minimizer term in the Eq. (13) as follows:

$$\inf_C \{F_1(C) + F_2(C)\} \approx 0 \approx F_1(C_0) + F_2(C_0) \quad (13)$$

By adding some regularization terms to the minimization equation above, the final energy functional $F(c_1, c_2, C)$ is defined by,

$$F(c_1, c_2, C) = \mu \cdot Length(C) + v \cdot Area(inside(C))$$

$$+ \lambda_1 \int_{inside(C)} |u_0(x,y) - c_1|^2 dxdy$$

$$+ \lambda_2 \int_{outside(C)} |u_0(x,y)$$

$$- c_2|^2 dxdy \quad (14)$$

where $\mu \geq 0$, $v \geq 0$ and $\lambda_1, \lambda_2 > 0$ are the fixed parameters. Therefore, the minimization problem proposed by the Chan-Vese method is as follows,

$$\inf_{c_1, c_2, C} F(c_1, c_2, C). \quad (15)$$

The problem can be redefined in a level set form where $C \subset \Omega$ represented by $\phi: \Omega \to \mathbb{R}$ as;

$$\begin{cases} c = \partial\omega = \{(x,y) \in \Omega : \phi(x,y) = 0\} \\ inside(C) = \omega = \{(x,y) \in \Omega : \phi > 0\} \\ outside(C) = \Omega \backslash \bar{\omega} = \{(x,y) \in \Omega : \phi < 0\} \end{cases} \quad (16)$$

The evolution of curve, $C$ in the above level of set functions of Eq.(16) is illustrated in Figure 13 which also reflects the relation between Chan-Vese [33] and Kass et al. [24] methods. Using the Heaviside function, $H$ in Eq.(17) and Dirac measure, $\delta_0$ in Eq. (18) the energy function can be rewritten as Eq.(19),

$$H(z) = \begin{cases} 1, & z \geq 0 \\ 0, & z < 0 \end{cases} \quad (17)$$

$$\delta_0(x) = \frac{d}{dz} H(z) \quad (18)$$

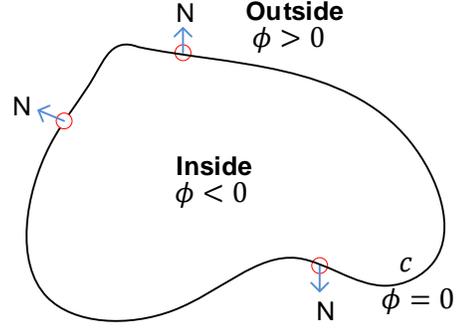

Figure 13: The propagation of curve C={(x,y):ϕ(x,y)=} in normal direction.

$$F(c_1, c_2, \phi) = \mu \int_\Omega \delta(\phi(x,y)) |\nabla\phi(x,y)| dxdy$$

$$+ v \int_\Omega H(\phi(x,y)) dxdy$$

$$+ \lambda_1 \int_\Omega |u_0(x,y)$$

$$- c_1|^2 H(\phi(x,y)) dxdy$$

$$+ \lambda_2 \int_\Omega |u_0(x,y) - c_2|^2 (1$$

$$- H(\phi(x,y))) dxdy \quad (19)$$

The variables $c_1$ and $c_2$ are fixed, and the energy function in Eq.(19) is minimized with respect to $\phi$ which concludes the Euler-Lagrange equation for $\phi$. The descent direction is chosen as an artificial time $t$ where the partial differentiation equation is expressed as follows;

$$\therefore \frac{\partial\phi}{\partial t} = \delta(\phi)\left[\mu \, div\left(\frac{\nabla\phi}{|\nabla\phi|}\right) - v - \lambda_1(u_0 - c_1)^2\right.$$

$$\left. + \lambda_2(u_0 - c_2)^2\right] = 0 \, in \, (0,\infty) \, X \, \Omega,$$

$$\phi(0, x, y) = \phi_0(x,y) \, in \, \Omega,$$

$$\frac{\delta(\phi)}{|\nabla\phi|} \frac{\partial\phi}{\partial\vec{n}} = 0 \, on \, \partial\Omega \quad (20)$$

where $\vec{n}$ represents the exterior normal to the boundary $\partial\Omega$, and $\frac{\partial\phi}{\partial\vec{n}}$ denotes the normal derivative of $\phi$ at the boundary. Lastly, Eq.(20) can converge to its solution by gradient descent method.

## B. Regularized Least-Square for $n^{th}$ Order Polynomial Fitting

Assume that we have $m$ points in 2D surface, $p_i = (x_i, y_i), i = 1 \dots m$, and we want to fit $n^{th}$ order polynomial where $m >> n$. The $n^{th}$ order polynomial, $P(x)$ can be expressed with $n+1$ coefficients as follows:

$$y = P(x) = \sum_{k=0}^{n} c_k x^k \tag{21}$$

One can turn this to a LS optimization problem by minimizing the total error,

$$c^{LS} = \min_{c \in R^{n+1}} \|y - P(x)\|^2$$
$$= \min_{c \in R^{n+1}} \left( \sum_{i=1}^{m} (y_i - P(x_i))^2 \right) \tag{22}$$

By defining the matrix, $A$ coefficient vector, x and output vector, b, this problem can be turned to a linear system as,

$$\begin{bmatrix} 1 & x_1 & x_1^2 & x_1^3 & \dots & x_1^n \\ 1 & x_2 & x_2^2 & x_2^3 & \dots & x_2^n \\ 1 & x_3 & x_3^2 & x_3^3 & \dots & x_3^n \\ \dots & \dots & \dots & \dots & \dots & \dots \\ 1 & x_m & x_m^2 & x_m^3 & \dots & x_m^n \end{bmatrix} \begin{bmatrix} c_0 \\ c_1 \\ c_2 \\ \dots \\ c_n \end{bmatrix} = \begin{bmatrix} y_1 \\ y_2 \\ y_3 \\ \dots \\ y_m \end{bmatrix} \tag{23}$$

or equivalently in a linear system equation,

$$A_{m \times (n+1)} \boldsymbol{c_{n+1}} = b_m \tag{24}$$

where $A$ is the $mx(n+1)$ matrix with the $n^{th}$ power of the x coordinates of the 2D points are the elements of the $(n+1)$-dimensional column vector, $\boldsymbol{c}$, of the polynomial coefficients, $(c_u)$ and $b$ is the $m$-dimensional column vector of y coordinates of the 2D points. For $m > n+1$, this is clearly an over-determined linear system, which means that there are more constraints (linear equations) than unknowns (parameters). In such systems there is no (exact) solution, one may only get a *unique* least-square (LS) solution as defined in Eq. (22) or equivalently, the LS solution of this equation, $c^{LS}$, can be expressed as follows:

$$c^{LS} = \min_{c \in R^{n+1}} \|b - Ac\|^2 = (A^T A)^{-1} A^T b \tag{25}$$

However, $A$ may not be even of full rank matrix, i.e., $rank(A) = r < n+1$, in which case, $A^T A$ will be *singular* and the inverse cannot be computed. To address this problem, we make use of the Singular Value Decomposition (SVD) of $A$ as follows,

$$A = U \Sigma V^T = \sum_{i=1}^{r} \sigma_i u_i v_i^T \tag{26}$$

where $U$ and $V$ are $mxm$ and $(n+1)x(n+1)$ orthogonal matrices which holds the eigenvectors of the square matrices, $AA^T$ and

$A^T A$, respectively, as the column vectors. The $mx(n+1)$ matrix, $\Sigma$, can be expressed as,

$$\Sigma = \begin{bmatrix} \sigma_1 & 0 & 0 & \dots & 0 & 0 & \dots & 0 \\ 0 & \sigma_2 & 0 & \dots & 0 & 0 & \dots & 0 \\ 0 & 0 & \sigma_3 & \dots & 0 & 0 & \dots & 0 \\ \dots & \dots & \dots & \dots & \dots & \dots & \dots & \dots \\ 0 & 0 & 0 & \dots & \sigma_r & 0 & \dots & 0 \\ 0 & 0 & 0 & \dots & 0 & 0 & \dots & 0 \\ \dots & \dots & \dots & \dots & \dots & \dots & \dots & \dots \\ 0 & 0 & 0 & 0 & 0 & 0 & \dots & 0 \end{bmatrix} \tag{27}$$

where $\sigma_1 > \sigma_2 > \dots > \sigma_r$ are the singular values or equivalently the eigenvalues of matrices, $A^T A$ and $AA^T$. This can yield the LS solution, $c^{LS}$, regardless whether or not $A$ is singular,

$$c^{LS} = V \Sigma^{-1} U^T b = \sum_{i=1}^{r} \frac{1}{\sigma_i} v_i u_i^T b \tag{28}$$

However, the LS solution, $c^{LS}$, can still yield large values, the so-called "explosion" of the LS solution, due to noisy values in matrix $A$ (the input = powers of x coordinates of the m points) or in vector $b$ (the output = y coordinates of the m points), or both. A crucial disadvantage here is that the smaller non-zero singular values will result in even larger explosion of $c^{LS}$. In order to prevent this, we shall *regularize* the LS solution by optimizing the LS error together with the magnitude of the LS solution as,

$$c^{RLS} = \min_{c \in R^{n+1}} (\|b - Ac\|^2 + \lambda^2 \|c\|^2) \tag{29}$$

where $\lambda$ is the regularization parameter. It is straightforward to show that this joint optimization can be expressed as,

$$c^{RLS} = \min_{c \in R^{n+1}} \left\| \binom{b}{0} - \binom{A}{\lambda I} c \right\|^2$$
$$= \min_{c \in R^{n+1}} \|b_\lambda - A_\lambda c\|^2 \tag{30}$$

where $A_\lambda$ is now an $(m+n+1) \times (n+1)$ full-rank matrix ($r=n+1$) and therefore, the LS solution over $A_\lambda$ can be obtained by using Eq. (25) as,

$$c^{LS}(\lambda) = \min_{c \in R^{n+1}} \|b_\lambda - A_\lambda c\|^2 = \left( A_\lambda^T A_\lambda \right)^{-1} A_\lambda^T b_\lambda$$
$$= (A^T A + \lambda^2 I)^{-1} A^T b \tag{31}$$

The $i^{th}$ eigenvector of $\left( A_\lambda^T A_\lambda \right)$ can be obtained by solving,

$$A_\lambda^T A_\lambda v_i = (A^T A + \lambda^2 I) v_i = (\sigma_i^2 + \lambda^2) v_i \tag{32}$$

So it is clear that matrix $A_\lambda^T A_\lambda$ has the same eigenvector, $v_i$, as matrix $A^T A$ but a larger eigenvalue, $(\sigma_i^2 + \lambda^2)$. Therefore, using the orthogonality of the eigenvectors, one can write the eigenvector decomposition of $A_\lambda^T A_\lambda$, and its inverse as follows:

$$A_\lambda^T A_\lambda = \sum_{j=1}^{n+1} (\sigma_j^2 + \lambda^2) v_j v_j^T = V \Lambda V^T$$

(33)

$$\left(A_\lambda^T A_\lambda\right)^{-1} = \sum_{j=1}^{n+1} \frac{1}{(\sigma_j^2 + \lambda^2)} v_j v_j^T = V \Lambda^{-1} V^T$$

Finally, using Eqs. (26) and (33) yields the regularized LS solution, $c^{RLS}$, expressed as,

$$c^{RLS} = c^{LS}(\lambda) = \left(A_\lambda^T A_\lambda\right)^{-1} A^T b$$

$$= \left(\sum_{j=1}^{n+1} \frac{1}{\sigma_j^2 + \lambda^2} v_j v_j^T\right)\left(\sum_{i=1}^{r} \sigma_i v_i u_i^T\right) b$$

$$= \left(\sum_{j=1}^{r} \frac{\sigma_j}{\sigma_j^2 + \lambda^2} v_j u_j^T b\right)$$

(34)

A direct comparison of Eq. (28) and (34) will reveal the fact that the regularized LS solution will no longer be effected from the noisy eigenvectors with very small eigenvalues, $\sigma_j$ since the when $\sigma_j \to 0$, also $\frac{\sigma_j}{\sigma_j^2 + \lambda^2} \to 0$ too with a reasonable choice of the regularization parameter, $\lambda$ (e.g., $\lambda \geq 0.1$).

### C. Computational Complexity Analysis

In this appendix, we shall first detail the computational complexities of the three main blocks: Active Contours (snakes) with $N$ vertices, $n^{th}$ order polynomial fitting and motion (displacement) computation of each segment.

#### 4) Active Contours

The active contour method used in the proposed method is Chan-Vese [33] which defines a minimization problem of an energy-based segmentation. The energy functional $F(c_1, c_2, C)$ is defined by [33] is as follows;

$$F(c_1, c_2, C) = \mu. Length(C) + v. Area(inside(C))$$
$$+ \lambda_1 \int_{inside(C)} |u_0(x, y) - c_1|^2 dx dy$$
$$+ \lambda_2 \int_{outside(C)} |u_0(x, y) - c_2|^2 dx dy$$

(35)

where $\mu \geq 0$, $v \geq 0$ and $\lambda_1, \lambda_2 > 0$ are the fixed parameters. The algorithm has a complexity of $O(mn)$ for each iteration (N = 300) where $m$ is the number of rows and, $n$ is the number of columns in the image. The sizes of the echo records in the dataset vary from 422×636 to 768×1024 pixels.

#### 5) $n^{th}$ Order Polynomial Fitting

The regularized LS solution for the $n^{th}$ order polynomial fitting to the m points in 2D is expressed in Eq. (22). The regularized LS solution, $c^{RLS}$ is expressed as,

$$c^{RLS} = c^{LS}(\lambda) = \left(A_\lambda^T A_\lambda\right)^{-1} A^T b$$

(36)

which can be expressed as a linear system as follows:

$$\left(A_\lambda^T A_\lambda\right) c^{RLS} = A^T b$$

(37)

where $A_\lambda$ is an $(m + n + 1) \times (n + 1)$ full-rank matrix and hence $\left(A_\lambda^T A_\lambda\right)$ is a $(n + 1) \times (n + 1)$ is a full-rank square matrix, $A^T b$ is a $n \times 1$ column vector. The solution of a linear system, $Ax=b$ where $A$ is an $n \times n$ full-rank matrix is in $O(n^3)$ in the worst case. This is negligible in this application since 4$^{th}$ order polynomials are used both for RPs and APs, and thus $n+1=5$ is the rank of the matrix $A$. Since we have $m$ points in 2D surface, $p_i = (x_i, y_i), i = 1..m$, and $m>>n$, the significant computational complexity will result from the matrix multiplication, $A_\lambda^T A_\lambda$, which is in $O(n^2(m+n))$ and matrix-vector multiplication, , which is in $O(nm)$. Therefore, both computational complexities are linearly proportional with the number of 2D points, $m$.

#### 6) Segment Motion Estimation and LVEF computation

This is the least computationally demanding block of all since the segment displacement can be approximated by averaging the point-wise distances as expressed in Eq.(3). With $L_2$ norm and $n_s = 5$ points, this requires only 5 summations and 10 differences, a total of 15 summation operation, which is negligible. This will be repeated by 6 segments for each frame in the echo, and for all frames so as to compute the maximum displacement of each segment. Similarly, the LVEF computation is even simpler than the segment-wise motion estimation since it requires only the counting (incrementing) the number of pixels inside the evolved snake.

#### 7) Computational Times

In the current un-parallelized and un-optimized MATLAB (version 2018a) implementation of the proposed method over a PC with 3.2GHz CPU and 16GB memory, the total execution time to process a cardiac cycle is about 36.9 seconds by 2 seconds per frame. The individual time share of each operation as illustrated in Figure 5 is given in Table 7. It is obvious that the majority of the computational complexity arises from the formation of the snake and then the formation of the both types of polynomials, RPs and APs.

Table 7: Execution time percentages of each block of the proposed method illustrated in Figure 5.

| Process | Time (%) |
|---------|----------|
| RPs | 17.45 |
| Snake | 78.15 |
| APs | 0.25 |
| LVEF | 0.19 |
| Disp. Ratio | 0.057 |